# Quality of Service Support on High Level Petri-Net Based Model for Dynamic Configuration of Web Service Composition

Sabri MTIBAA
LI3 Laboratory / University of Manouba
National School of Computer Sciences
2010 Manouba, Tunisia
Sabri.Mtibaa@gmail.com

Moncef TAGINA
LI3 Laboratory / University of Manouba
National School of Computer Sciences
2010 Manouba, Tunisia
Moncef.Tagina@ensi.rnu.tn

***Abstract*— Web services are widely used thanks to their features of universal interoperability between software assets, platform independent and loose-coupled. Web services composition is one of the most challenging topics in service computing area. In this paper, an approach based on High Level Petri-Net model as dynamic configuration schema of web services composition is proposed to achieve self adaptation to run-time environment and self management of composite web services. For composite service based applications, in addition to functional requirements, quality of service properties should be considered. This paper presents and proves some quality of service formulas in context of web service composition. Based on this model and the quality of service properties, a suitable configuration with optimal quality of service can be selected in dynamic way to reach the goal of automatic service composition. The correctness of the approach is proved by a simulation results and corresponding analysis.**

## I. INTRODUCTION

With the evolution of network and service infrastructure towards Service-Oriented Architectures (SOA), an important requirement for new application components is to present a high level interface that allows developers to use and re-use such components known as Web Services (WS) in new applications. WS, as they are easily accessible from any point of internet, are suitable to build rapidly on-demand applications.

From point of view of their internal complexity, WS can be divided into two categories: *elementary* WS and *composite* WS.

Elementary WS offer a basic service, like simple libraries and contain a low level of data transformation, for example, translation services are elementary WS. In the contrary, composite WS has more complex pattern and more powerful function. Such a composite service, resulting of the composition of several processes logically assembled, can be called an orchestrated service. For instance, converged services which are involving combined reusable components based on network capabilities such as calling services, messaging services and internet services.

Although Web services technologies and BPEL (Business Process Execution Language)-based orchestration engines are powerful tools for developing blended applications, they are just examples of the tools available in this space. These tools are well suited for developing capabilities that are abstracted at a fairly high level and have less stringent latency and real-time requirements. For example, a service sending a time-sensitive message needs to consider all aspects of delays and latencies involved in the overall service flow. If a location-based advert reaches the person after the fact, the advert loses its meaning.

In Web services composition context, there often exist a number of alternative component systems which have the same functionality but differ in QoS (quality of service). In the recent academia researches [1, 2], the problem of web service selection was deeply studied through performance evaluation and estimation methodologies. For this purpose, some methods and tools to capture and analyze the performance of WS have been developed [3]. In general, the proposed approaches for evaluation of quality of service capabilities for WS are quite different each from the other. Each one focuses on a different set of QoS metrics and can be applied at run-time or system design time.

In the actual applications, a functional relationships dependency between services exists. A service selection is very probably influencing the next service selection and consequently the Web services execution flow. The web services composition must have the ability of dynamic reconfiguration in order to adapt to the change of run-time environment [4].

Reconfigurability is an important feature of self-adaptive system, especially high-assurance system. San-Yih Hwang et al. formulated the dynamic WS selection procedure in a dynamic environment that is failure prone. They proposed FSM (Finite State Machine) usage to invoke operations of service in an order [4]. The Web services are selected dynamically at run time in their work. In [5], a composite configuration schema is modeled by Petri net. These works have made a good job in modeling of service dependent relationship. In self-configuring approaches, like those presented in [6] and [7], service selection is performed by searching for an optimal configuration of components based upon the initial constraints.

However these solutions present a complexity in the integration with available service orchestration engines. A main reason is the orchestration process of composite service is not considered in their models. Expected benefits are a better productivity, the ability to be much more reactive to requests. Without considering the orchestration process, it will be difficult to well evaluate whether the user's QoS requirements are satisfied or not.

Most of theories such as finite state machine, Pi-calculus and Petri nets have been used for description of web services composition. Their main concern is model mapping: modeling through translating composition plan or language into model [8, 9]. In [10], an approach oriented towards QoS assurance for the evaluation of different design alternatives using a discrete-event modeling is presented. However, it cannot satisfy the requirements of adaptive and automatic system management. In order to construct a self-adaptive composite service, different techniques such as simulated annealing, stochastic Petri net [11] are presented to accelerate service global selection.

In this paper, we focus on this paper on modeling composite service configuration schema and service orchestration process. The dependency relationship and the possible orchestration processes are reflected by a high level Petri net presented in this study. The support of QoS attributes and properties calculated under different configurations help to select dynamically the best and optimal configuration. This final configuration and orchestration process can be directly used by the current service engine.

The remainder of this paper is organized as follows. In Section II, we present a web services overview. Section III introduces the dynamic configuration model using hierarchical Petri net. In Section IV, we present our QoS calculation method applied on the Petri-net based model. A selection of dynamic configuration is described in Section V. Section VI presents a case study to illustrate our work. Finally, we conclude in section VII.

## II. Web Services Overview

SOA is architecture that functions are defined as WSs. According to [10], WSs are self-contained, modular applications that can be described, published, located, and invoked over the network, generally, the World Wide Web. The SOA is described through three different roles: service provider, service requester and service registry.

The key idea of SOA is the following: a service provider publishes services in a service registry [12]. The service requester searches for a service in the registry. He finds one or more by browsing or querying the registry. The service requester uses the service description to bind service. These ideas are shown in Fig. 1.

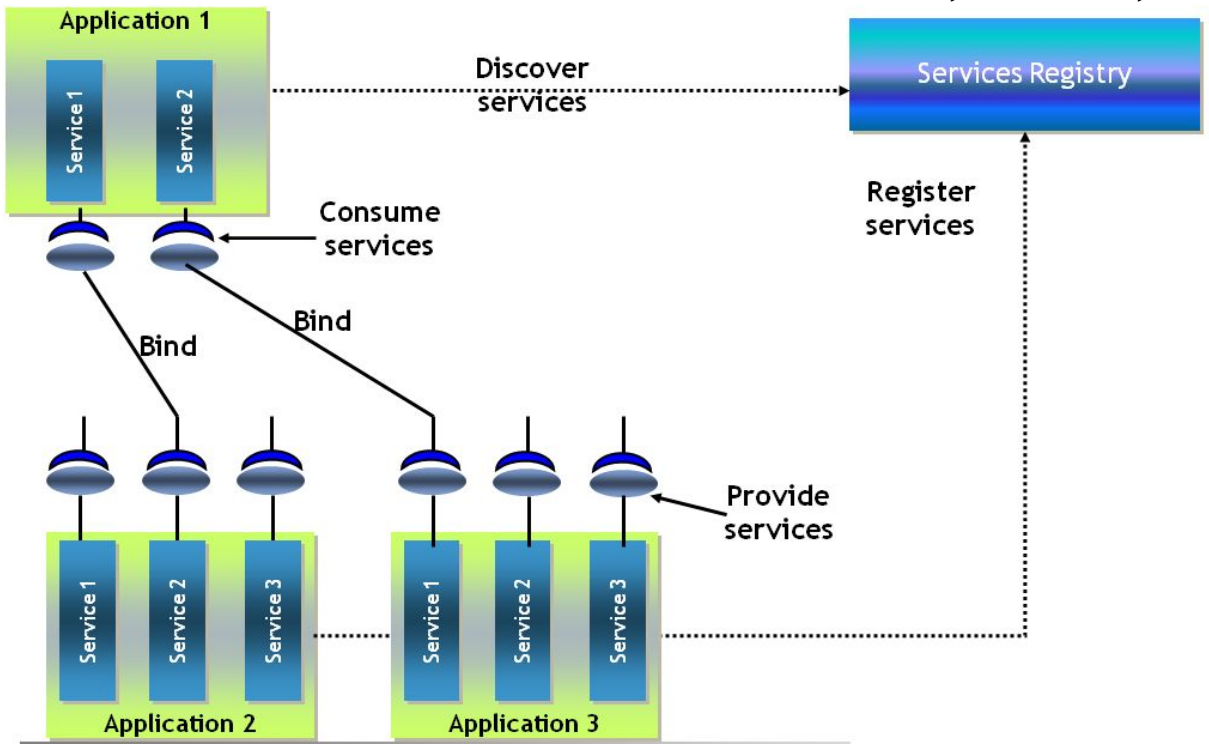

Figure 1. Service-Oriented Architecture

The composition mechanism leads us from the elementary components to the final new service.

Orchestration is the term used to describe the creation of a "business process" (or a workflow) using Web services. A business process is an aggregation of services whose the operations, i.e. the processes, are logically linked together in order to reach a given objective [1].

Aggregating services to build an added-value service have many solutions depending on the chosen environment. For Web services, the orchestration is usually expressed with a specific language like BPEL that describes the interactions between the services.

A business process is deployed itself as a service, so it can be used by other processes. A business process language describes the behaviour of business processes based on Web services, i.e.:

- Control flow (sequences, loops, conditions, parallelism …)
- Variables, exceptions, timeout management.

SOA is known to bring many advantages in software development, management and deployment. A product is a set of SOA services focused on solving on a business problem (see Fig. 2). The focus is on e.g. the following points [13]:

- **Modularity/re-usability**: Individual components of a solution can be managed independently of each other, allowing components to follow different release cycles and frequencies of release, with minimal regression across them.
- **Maintainability/evolution**: Components have "hard edge" APIs, so can be substituted as required to address issues such as performance, scalability and stability on a more granular, case-by-case basis.
- **Loosely coupled**: Reduces the risk of "domino effect" total service outages, as components and processes are less tightly bound to one another.
- **Flexibility**: Enables more flexible distribution and placement of components across multiple system resources in n-tier service model, whilst not imposing a static nor a rigid deployment model.

- **Scalability**: Service capacity requirements can be managed in an easy, sustainable, repeatable and consistent manner.

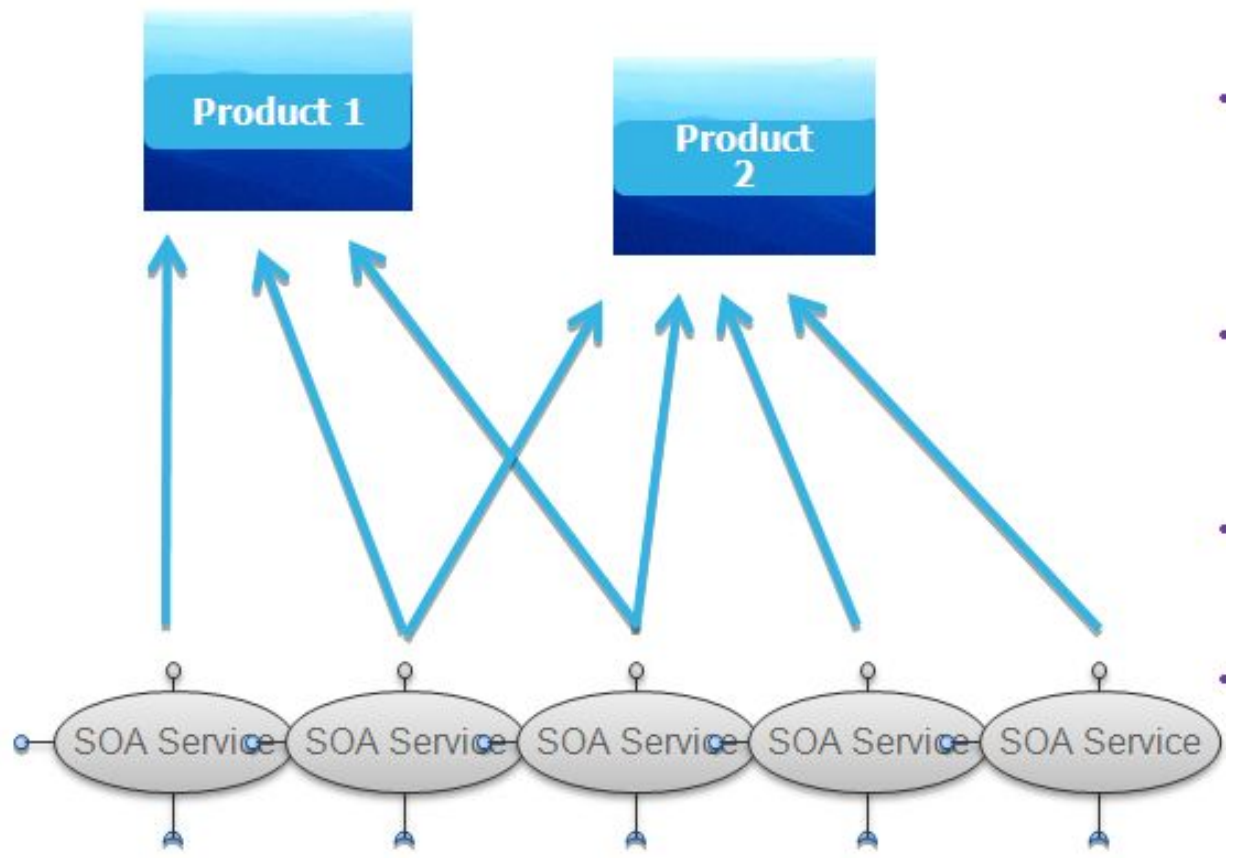

Figure 2. Products as a composition of SOA services

## III. DYNAMIC CONFIGURATION MODEL

In our research, we chose to adopt Petri nets due to its combination of (1) rich computational semantics, (2) ability to formally model systems especially with properties: concurrent, asynchronous, distributed, parallel, nondeterministic and stochastic, and (3) availability of graphical simulation tools [14].

Petri nets also have natural representation of changes and concurrency, which can be used to establish a distributed and executable operational semantics of Web services [15]. In addition, Petri nets can address offline analysis tasks such as Web services static composition, as well as online execution tasks such as deadlock determination and resource satisfaction. Furthermore, Petri nets possess natural way of addressing resource sharing and transportation, which is imperative for the Web services paradigm.

### *A. Model Definition*

**Definition 1:** The algebraic structure of Hierarchical Timed Predicate Petri Net named HTPPN = (P, T, F, N, K, W, TC, TD, $Q_c$ ,$P_i$, $P_0$) if the following conditions hold:

- $P \cap T = \varnothing$ , $P \cup T \neq \varnothing$, P is sets of places
- T= $T_R \cup T_d \cup T_c$, $T_R \cap T_d = \varnothing$, $T_d \cap T_c = \varnothing$, $T_R \cap T_c = \varnothing$. $T_d$ is a set of dummy transitions, $T_c$ is a set of concrete transitions, $T_R$ is a set of refinable transitions. For $t_d \in T_d$, $t_d$ can be associated with a choice probability. For $t_R \in T_R$, $t_R$ is a HTPPN with unique input transition $t_i$ and unique output transition $t_0$.
- $F \subseteq (P \times T) \cup (T \times P)$ , F is a finite set of arcs
- N: T $\rightarrow$ Name $\cup$ {$\phi$}, Name represents the name of web services composition. $\phi$ represents that the operation of web service is not actual execution but structural simulation.
- K: P $\rightarrow$ $N^+$ is a capability function. N+ is the set of positive integers.
- W: F $\rightarrow$ $N^+$ is a weight function.
- $Q_c$: $T_c \rightarrow R^+$ is a QoS function. $R^+$ is a set of positive real numbers.
- $P_i \in P$
- $P_0 \in P$

T is a finite set of transitions which represents the activity of web service. F is called the web services action flow.

A marking in a HTPPN is a function M that maps every place into a natural number. $M_0$ is called the initial marking.

The execution model is defined with the following basic temporal types: time point, duration and interval constraints. They are defined as follows:

- TEB(j) is the sum of Message Delay Time and waiting time of the activity j of a process instance.
- [$TEB_{min}$(j),$TEB_{max}$(j)] denotes the time period during which the activity *j* are enabled after the intermediate preceding activity of the activity j finishes.
- [$TEC_{min}$(j),$TEC_{max}$(j)] denotes the time period during which the activity j can be executed after it is enabled.

TC = TC(p)$\vee$TC(t), where TC(p) is a set of all place time pairs and TC(p)={ [$TC_{min}$(p),$TC_{max}$(p)]$\in$Z x Z | $TC_{min}$(p) < $TC_{max}$(p) $\wedge$ p $\in$ P }, TC(t) is a set of all transitions time pairs and TC(t)={ [$TC_{min}$(t),$TC_{max}$(t)]$\in$ Z x Z | $TC_{min}$(t) < $TC_{max}$(t) $\wedge$ t $\in$ T }. TD$\in$Z is a set of time duration.

The model has unique input place $P_i$ and unique output place $P_0$.

- The model is also a Petri net, i.e., all of its arcs weights are 1.
- $|T_m| \geq 1$. The model contains at least one refinable transition.
- Each concrete transition is associated with a component service.
- Each refinable transition model is constructed by a deferred choice pattern.
- $M_0(P_i)$ = 1. □p □ P\ {$p_i$}, $M_0$(p) = 0

**Definition 2: (Transition firing/ Transition firing duration)** For t $\in$ $T_c$, if TC (t) $\in$ [$TC_{min}$(t),$TC_{max}$(t)], then a transition t can fire at least after $TC_{min}$ unit time intervals if it is enabled at marking M; during the period, transition t must fire at most after $TC_{max}$(t) unit time intervals if there is no transition enable, which may change marking and make the transition t unreachable.

**Definition 3: (Transition enabled interval/ Token arrival)** The interval [$TC_{min}$(P),$TC_{max}$(P)] presents the time period during which $P_i$ succeeding transitions are enabled after a token arrives at a place $P_i$.

**Definition 4: (Time consistency of composite WSs )**
As to a service, the perform execution time span is mapped to earliest and latest enable time of transition as ($TEB_{min}(j)\ldots TEB_{max}(j)$), while the execution time is denoted by firing duration of transition as *TE(j)*. Then a transition is considered as schedulable if it is candidate to fire and can finish its firing successfully. i.e., ($TEB_{max}(j)$ - $TEB_{min}(j) > 0$).

A marking $M_n$, is said to be reachable in TPPN modeling if there is a firing sequence ($M_0\ t_1 M_1\ .\ .\ .\ M_j\ t_j\ .\ .\ .\ M_n\ t_n$) that transforms $M_0$ to $M_n$. In a TPPN, if any transition is schedulable, we show that service can be successfully within time constraints.

## *B. Model Description*

The HTPPN is a model representing the service selection as well as the orchestration process. Places are designed for the system states and the execution conditions.
A web service behavior is basically or partially ordered set of activities. For instance, Figure 3 shows typical Petri Nets model that represents the relationship between places (i.e. state of the service) and transition (i.e. a given activity).

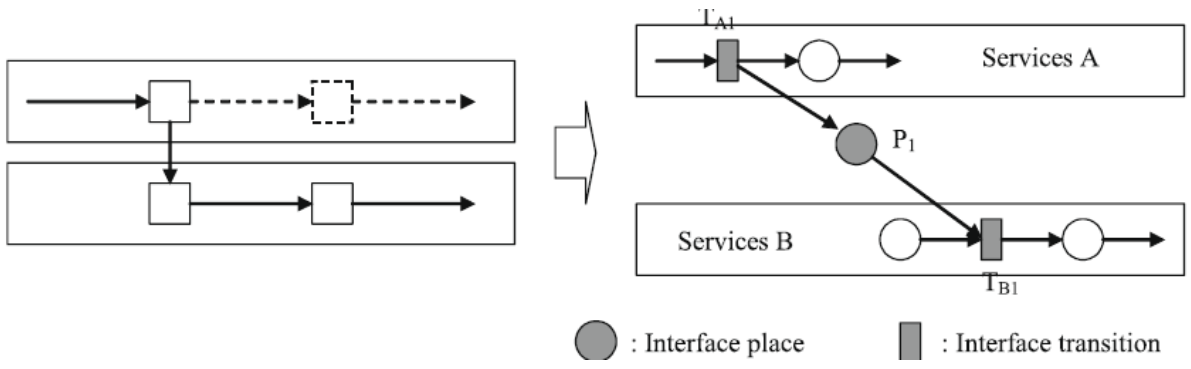

Figure 3. Model of a simple chained flow

The logical control activities in the composition process (pre-condition and post-condition) are represented by dummy transitions. No component service is associated to a dummy transition. The activities of component services are represented by concrete transitions. Only one fixed component service is associated to a concrete transition.
There are several configurations in a HTPPN. A feasible configuration can be get after one selection branch is retained and others are eliminated in every refinable transition. Since different configuration is candidate to include several number of component services, the corresponding orchestration process can also be different. In this work, HTPPN reflects not only the possible service selection but also the corresponding orchestration process.

Service selection means the process to associate every activity of composite service to a component service. According to whether a service selection is related to another, the service selections can be classified into two families, free selection and restricted selection.

Figure 4 and 5 depict the two kinds of selections modeled by using *split-join concurrency pattern* based Petri net.

In Figure 4, every selection branch only includes one transition. Therefore, the transitions can be selected freely. In contrary, every selection in Figure 5 branch includes a set of transitions. The service dependent relationships are reflected by the flow pattern of branch. The transitions in a branch must be selected as a whole. For example, the set {$T_1$, $T_3$} or {$T_2$, $T_4$} is selected.

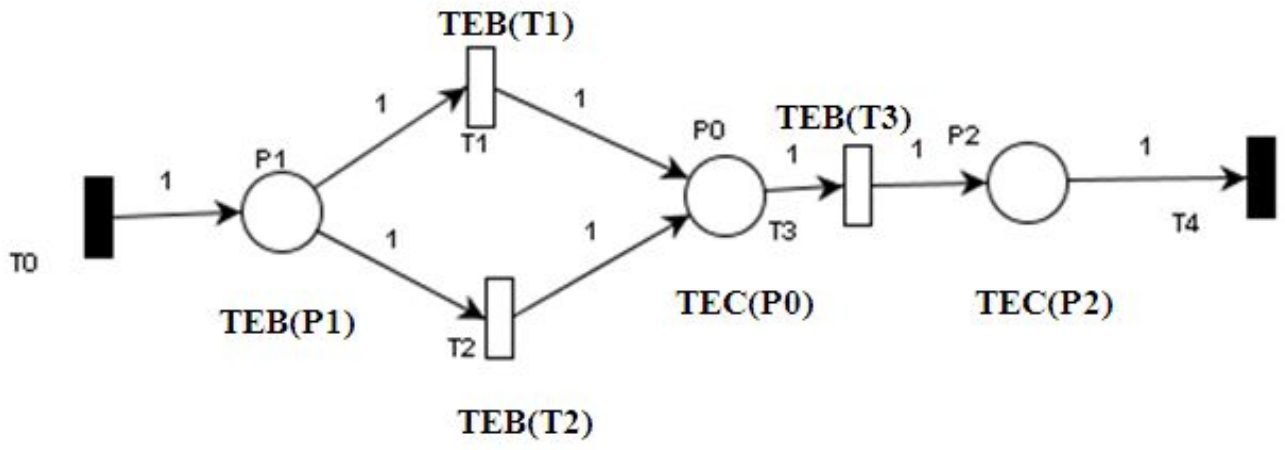

Figure 4. A modeling example of dynamic configuration using HTPPN (without dependent relationship)

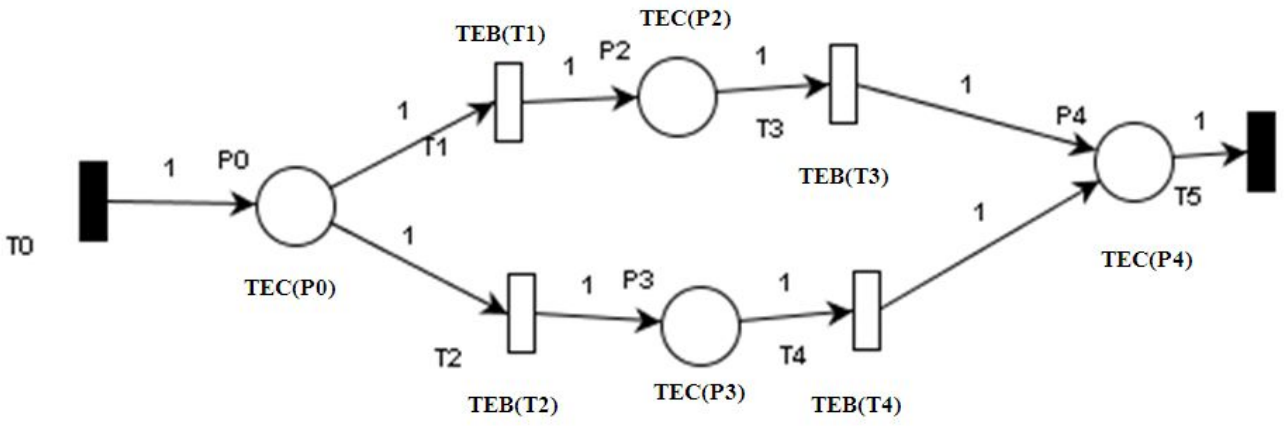

Figure 5. Example of HTPPN model representing the service selection with dependent relationship

The activities of service selection are encapsulated in the refinable transitions. The refinable transition indicates that an abstract sub-function of composite service can be accomplished by a number of optional component services or service modules. Each branch of *split-join concurrency pattern* in a refinable transition represents a service selection (see Figure 6). The QoS attribute is associated to concrete transitions.

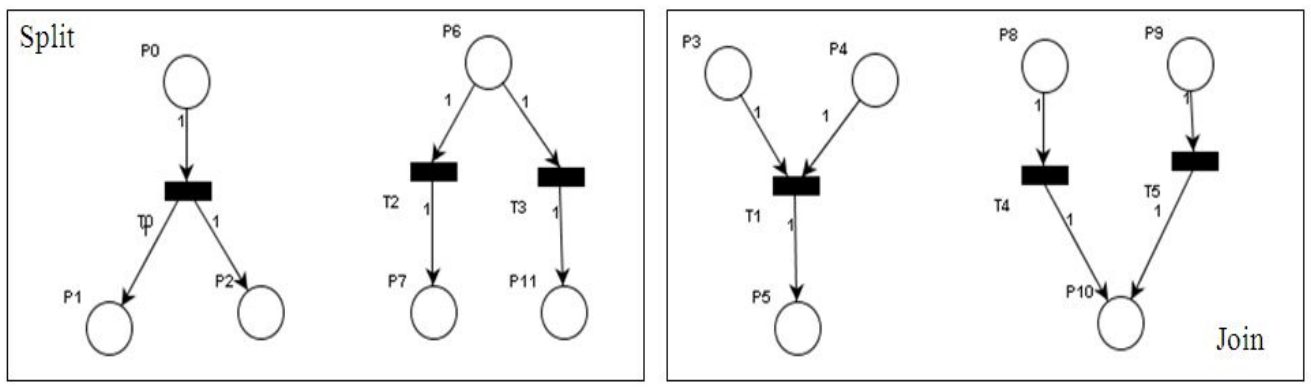

Figure 6. A split-join concurrency pattern

## IV. QoS Calculation Method

In this section, a QoS calculation method of composite service is presented at first. Some QoS attributes, such as response time, throughput, cost, reliability, availability, security, accessibility have been exposed to evaluate a Web service. Each component service may have several QoS attributes. QoS attributes such as throughput, reliability, availability, security and accessibility are better if larger value. But for attributes like response time and cost, the smaller is the value, the better is the QoS.

In this study, a simplification of the model is made that consider smaller value as better QoS. For larger-is-better attributes, their values will be changed automatically to negative number. If a component service is unavailable or disabled the QoS value of it is set to + ∞.

Because a selection branch in a refinable transition according to split-join pattern must be selected as a whole, the QoS of the branch needs to be evaluated. Different branch patterns need to be considered. HTPPN is usually comprised of four kinds of basic Petri net patterns. The α is a choice probability associated to dummy transition. To different QoS attributes, the QoS calculation formulas of four kinds of patterns are different.

We propose some feasible estimate formulas to calculate the QoS of basic model patterns. R(t), C(t), A(t) and T(t) are respectively response time, configuration cost, availability (or reliability) and throughput of component service. The configuration cost indicates the charge being paid for the third-party component services. It can be simply the summing up of all the costs of component services. The formulas are easy to be understood and verified, except QoS formulas of loop pattern presented in [9]).

### A. QoS Calculation Formulas

We figure out some useful patterns defined in Workflow Management Coalition (WFMC):

- *Sequential pattern :*

The sequence construct allows the definition of a collection of activities to be performed sequentially in lexical order (see Figure 7). A sequence activity contains one or more activities that are performed sequentially.

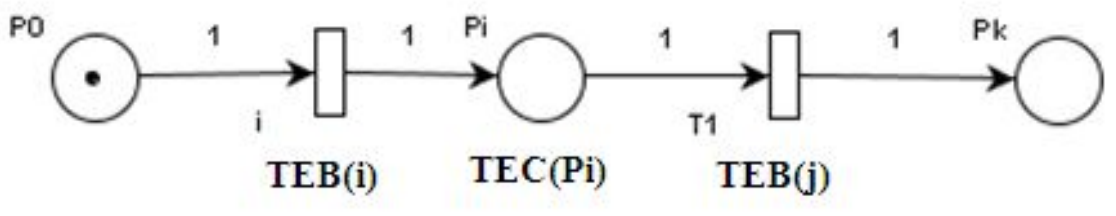

Figure 7 Petri-Nets based sequential pattern

The processes *i* and *j* execute in sequential order, QoS calculation formulas are:

$$R(i \circ j) = TEB(i) + TEB(j) + TEC(P_i)$$
$$C(i \circ j) = C(i) + C(j)$$
$$A(i \circ j) = R(i) \times R(j)$$
$$T(i \circ j) = Min(T(i), T(j))$$

- *Parallel pattern :*

The flow construct allows to two or more activities to be executed in parallel, giving rise to multiple threads of control. Transition $T_0$ (see Figure 8) is a point where a single thread of control splits into two or more threads that are executed in parallel, allowing multiple activities to be executed simultaneously.

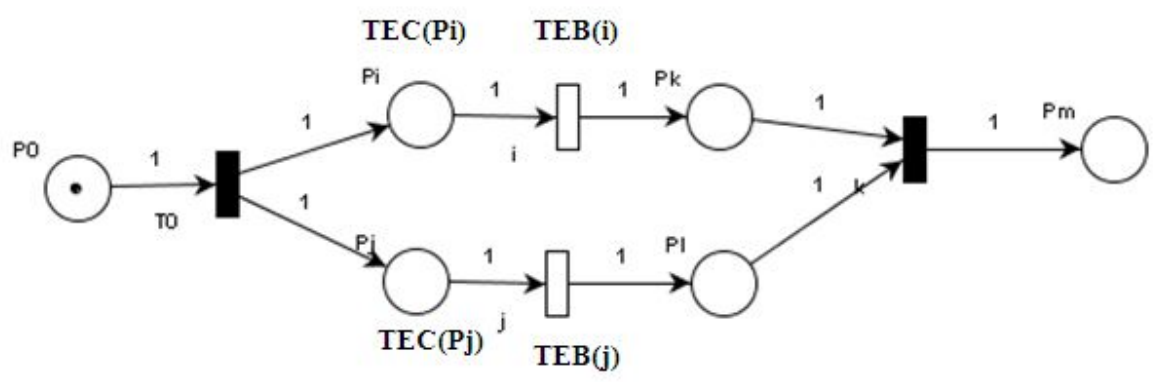

Figure 8. Petri-Net based parallel pattern

The processes *i* and *j* execute concurrently, QoS calculation formulas are:

$$R(i \| j) = Max(TEB(i), TEB(j))$$
$$C(i \| j) = C(i) + C(j)$$
$$A(i \| j) = R(i) \times R(j)$$
$$T(i \| j) = Min(T(i), T(j))$$

- *Conditional pattern:*

The flow construct is used to select exactly one branch of execution from a set of choices. It supports conditional routing between activities. Place $P_i$ (see Figure 9) is a point where a single thread of control makes a decision upon which branch to take when encountered with multiple alternative activity branches.

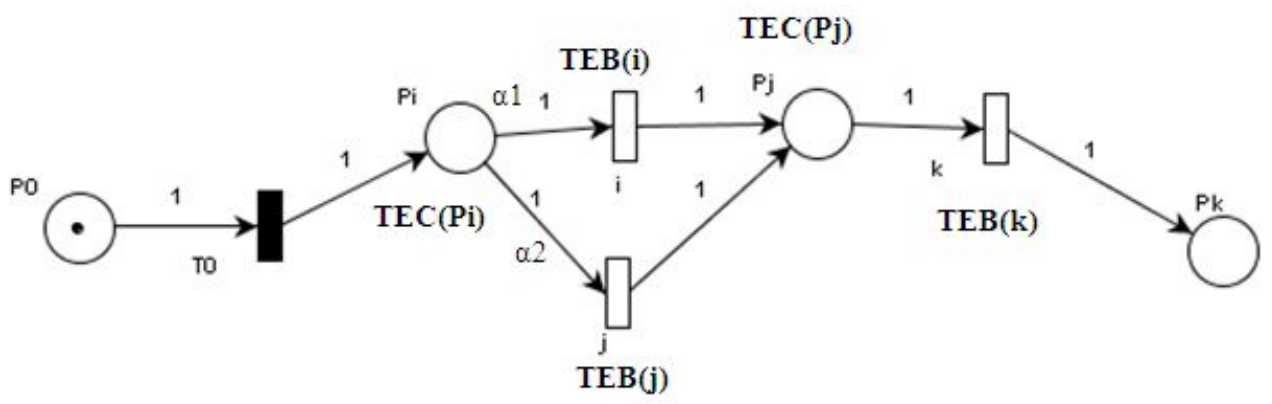

Figure 9. Petri-Net based conditional pattern

Suppose that processes *i* and *j* execute alternatively, QoS calculation formulas are:

$$R(i \vee j) = Max(TEB(j) + TEC(P_j), TEB(i) + TEC(P_j))$$

$$C(i \vee j) = C(i) + C(j)$$
$$A(i \vee j) = \alpha_1 R(i) + \alpha_2 R(j)$$
$$T(i \vee j) = \alpha_1 T(i) + \alpha_2 T(j)$$

- *Loop pattern :*

The loop construct (Figure 9) is used to indicate that an activity is to be repeated until a certain success criteria has been met. A while activity supports repeated performance of an activity in a structured loop, that is, a loop with one entry and one exit point.

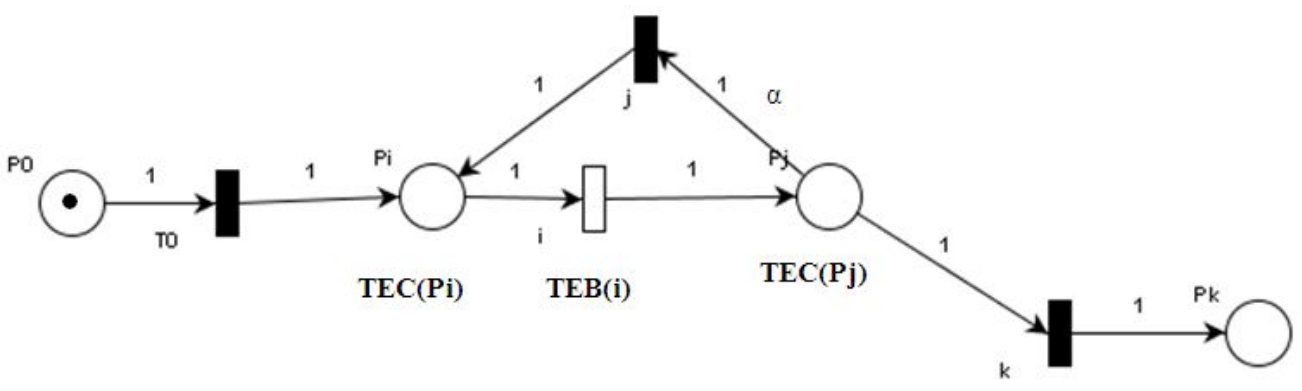

Figure 10. Petri-Net based Loop pattern

If we suppose that process *i* execute k times, QoS calculation formulas are:

$$R(k \times i) = (\sum_{j=1}^{k} TEB_{\min}(j), \sum_{j=1}^{k} TEB_{\max}(j))$$
$$C(k \times i) = C(i) + C(j)$$
$$R(k \times i) = \frac{(1-\alpha)R(i)}{1-\alpha R(i)R(j)}$$
$$T(k \times i) = (1-\alpha)t(i) + \alpha Min(T(i), T(j))$$

## B. Dynamic Configuration Model

A service orchestration model with static configuration can be transformed to one with dynamic configuration. For example, in Figure 11, if there is a number of component services with identical function, either of which can accomplishes the function of $T_{41}$ or $T_{42}$, then $T_4$ is transformed to a refinable transition (modeled as a selection pattern shown in Figure 4).

Some important modeling trips need to be highlighted and detailed:

• In Figure 11, one selection branch depicted by refinable transition $T_6$ can include different flow pattern and different quantities of transitions in contrast with another, as long as each branch can accomplish the same function. This enables the model to reflect the various service orchestration processes.

•The refinable transitions can also exist in hierarchical manner, i.e., may exist other refinable transitions in a refinable transition. A complex composite service configuration schema can be compactly reflected by the N-hierarchical model.

•The service configuration schema and optional orchestration processes are synthetically reflected in a HTPPN. But, only one selection branch in every refinable transition should be kept in the final model.

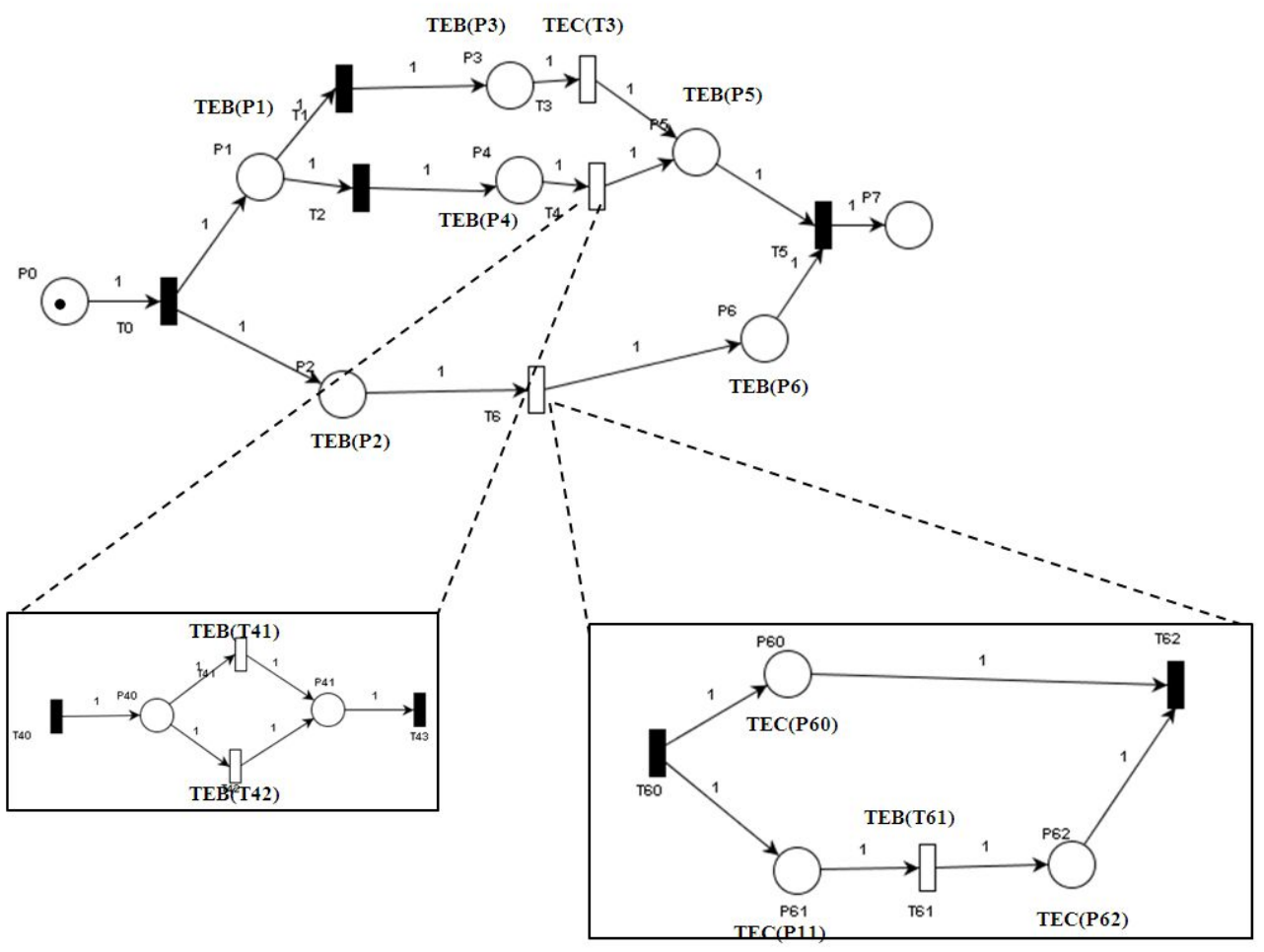

Figure 11. Example of HTPPN model representing a dynamic configuration

## V. Selection of Dynamic Configuration

An optimal configuration selection algorithm is designed to select the configuration with best QoS from the service configuration schema as shown in Figure 12.

The algorithm, presented in Figure 12, considers HTPPN as input and provides as output an optimal QoS configuration. It's an iterative algorithm that allows considering different levels through the hierarchical model.

In each iteration, the current selection branch is picked defined and the *caclulateQosForEachBranch* method is called. This method calculates the QoS attributes using the formulas presented according to Petri-nets patterns. Then *selectBranchWithMinimalQoS* method is executed in order to select the best configuration having the less attributes values.

The *EntryModel* is the main model in HTPPN. This model is the first level in the hierarchy. Each branch represents a selection alternative that will be considered in the orchestration.

```
optimalConfigurationSelection (input : HTPPN)
 output : optimal QoS configuration
 {
   S ← ∅  # S is stack of refinable transitions
    For t ∈ T_R(HTPPN) do t.selected=false done
    For t ∈ T_R(EntryModel) do push(t,S) done
    while S != ∅
     T_i=pop(S)
     if |T_R(T_i)| t ∈ T_R(HTPPN) || ( t ∈ T_R, (T_i,t.selected=true))
       then caclulateQosForEachBranch()
              selectBranchWithMinimalQoS()
            T_i.selected=true
       else push(T_i,,S)
              For t ∈ T_R(T_i) do push(t,S)done
     end if
     end while
 }
```

Figure 12. Optimal selection algorithm

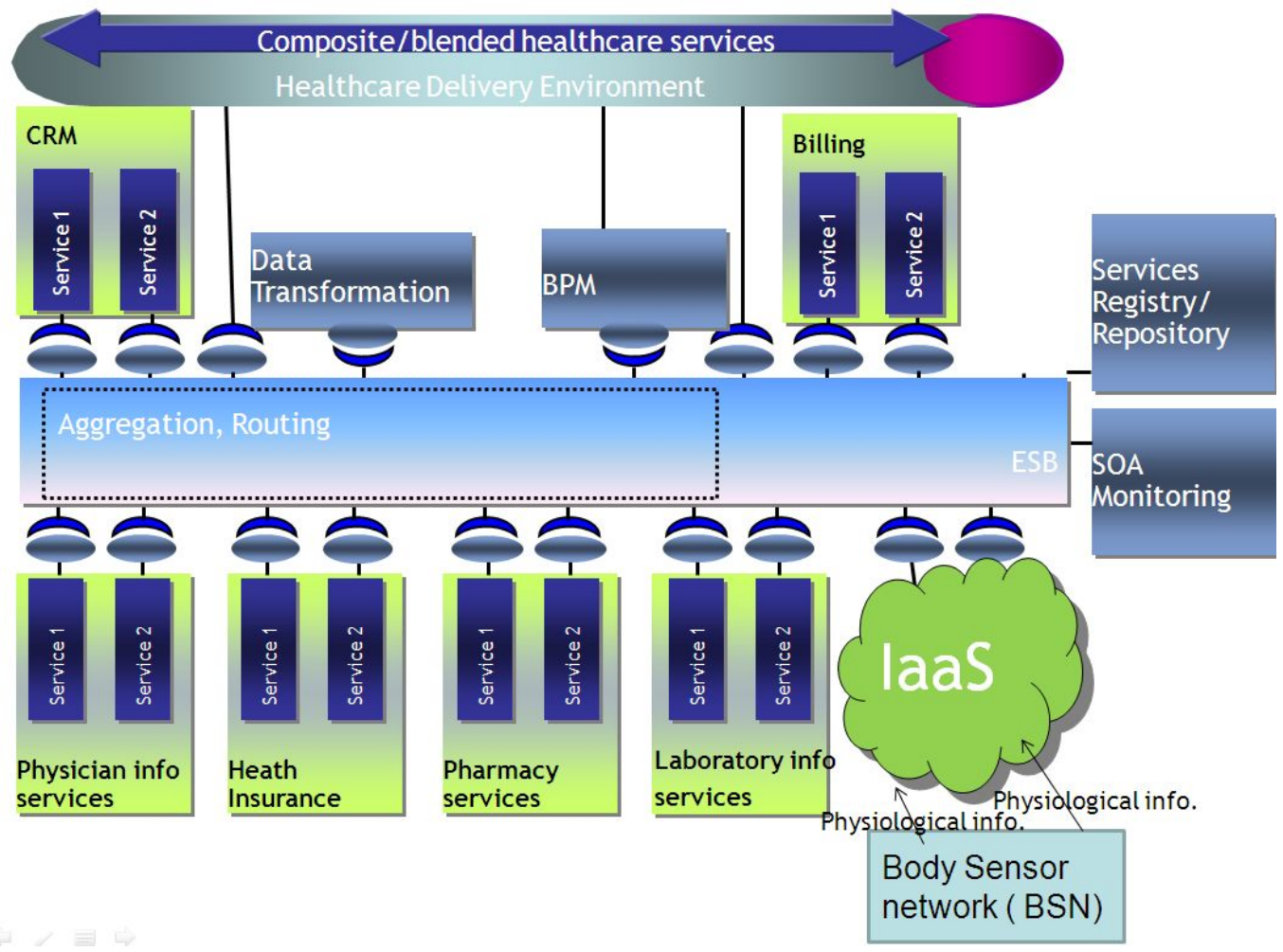

Figure 14. HSP architecture

## VI. CASE STYDY

### A. Description

The Healthcare Service Platform (HSP) presented in [16] in focuses on the delivery of healthcare services. It is an end-to-end reference architecture that focuses on meeting the needs of citizens, patients and professionals. Its architectural diagram is given in Figure 14.

We distinguish three main components, i.e. body sensor networks (BSN), IaaS cloud, healthcare delivery environment.

- *BSN*: according to circumstances and personalized needs, appropriate health information collection terminals (i.e. sensors) are configured for different individuals. BSN is used to provide long term and continuous monitoring of patients under their natural physiological states. It performs the multi-mode acquisition, integration and real-time transmission of personal heath information anywhere.
- *IaaS cloud*: this component achieves the rapid storage, management, retrieval, and analysis of massive heath data. It mainly includes *Electronic Medical Record* (EMR) repository. It considers also personal health data acquired from BSN.
- *Healthcare delivery environment*: it includes a personal health information management system. It replaces expensive in-patient acute care with preventative, chronic care, offers disease management and remote patient monitoring and ensures health education/wellness programs.

In HSP, we adopt the design idea of SOA and Web service technology for its design and implementation. The majority of its functional modules are developed and packaged in the form of services. Here, we overview some of them as follows [17].

- *PhyInfoService*: this service can acquire some general physiological signals such as body temperature, blood pressure, and saturation of blood oxygen, electrocardiogram, and some special physiological signals according to different sensor deployment for different users. User’s ID number is required.
- *EnvInfoService*: for a unique ID number, this service can acquire temperature, humidity, air pressure and other environmental information for this user.
- *SubjFeelAcqService:* it can acquire the user subjective feelings, food intake, etc., and the information is often provided by the user from the terminal.
- *TempInfoService:* it can return the external temperature in the patient’s environment.
- *HealthGuideAssService:* this service can assess the knowledge of the patient’s health risk based on specific questions.
- *EMRService:* this service can output the user’s medical history information.
- *GeoInfoService*: it can return the user’s location.
- *EmerAlarmService*: it can raise an alarm to the user in case of illness.
- *HealthGuideService*: it can provide the patient with preventive measures especially items that need attention.
- *RealTimeWarmService*: it can warm the patient about the signs of certain disease.

We present the HTPPN model of a healthcare scenario. We highlight that the transition with H as symbol represents a refinable transition.

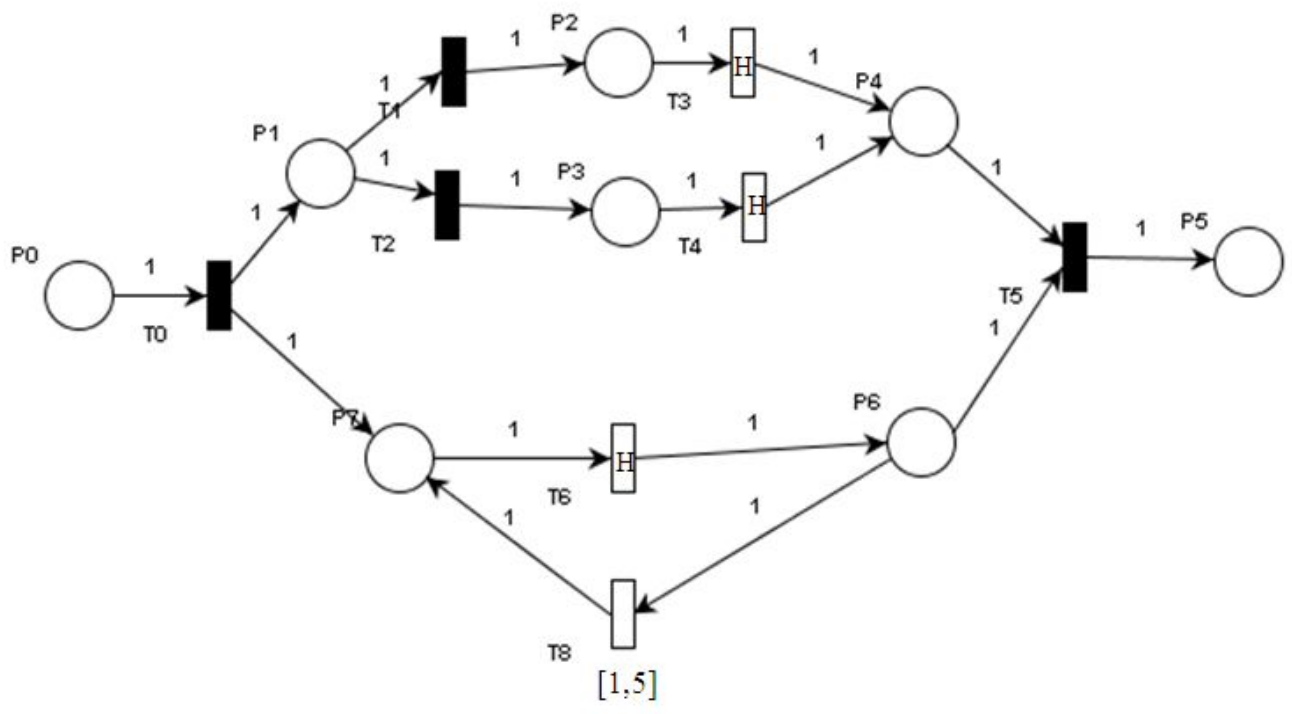

Figure 15. An example of HTPPN modeling a Healthcare scenario

In our model, HTPPN–HSP mapping is as described in Table I. HTPPN represents the EntryModel (see Figure 15).

TABLE I. HTPPN-HSP MAPPING

| Semantic Map | |
|---|---|
| ***T3*** | ***ThirdService*** |
| ***T4*** | ***HealthCareService*** |
| ***T6*** | ***MedicalAnalysisService*** |
| ***T8*** | ***AssesmentService*** |

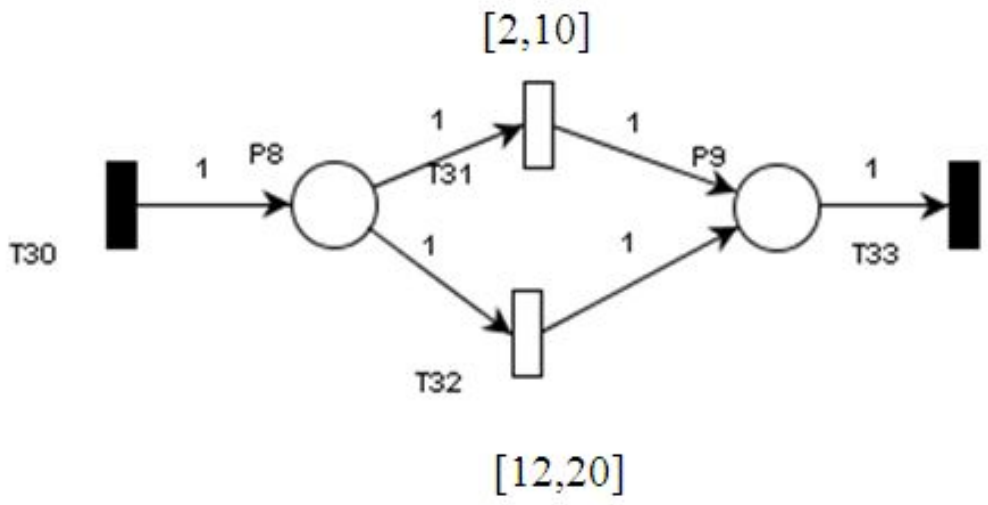

Figure 16. The model of refinable transition $T_3$

Table II presents $T_3$–HSP mapping. It's associated with refinable transition $T_3$ (see Figure 16).

TABLE II. $T_3$-HSP MAPPING

| Semantic Map | |
|---|---|
| ***T31*** | ***FinancialService*** |
| ***T32*** | ***InsuranceService*** |

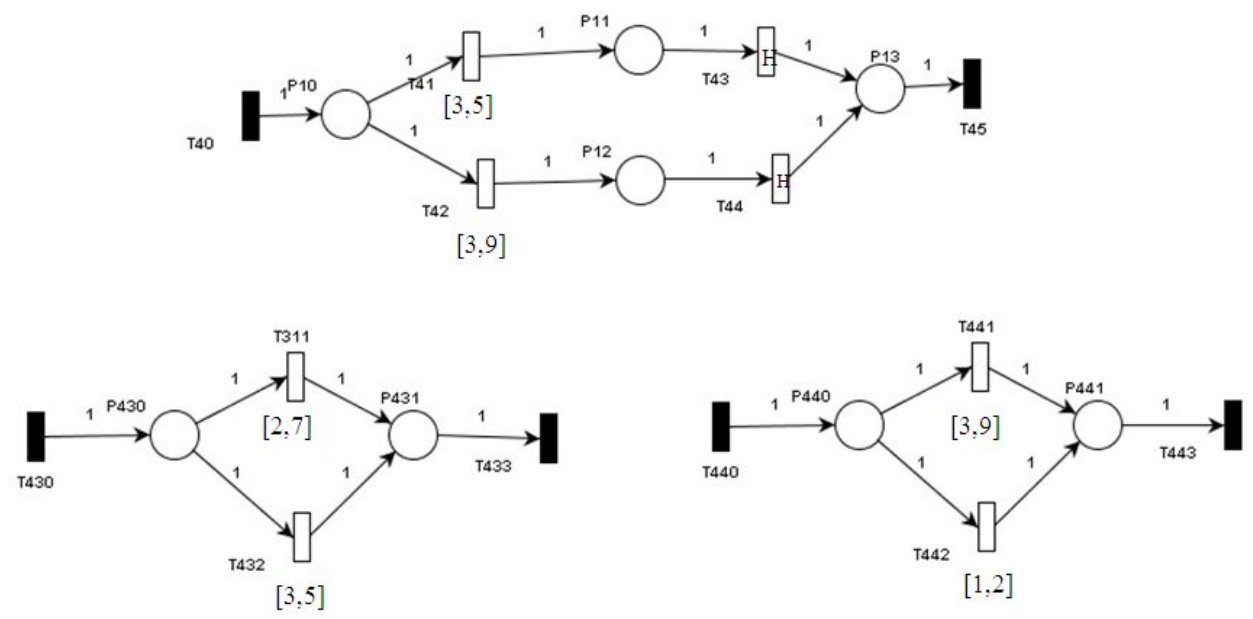

Figure 17. The model of refinable transition $T_4$

Table III depicts $T_4$–HSP mapping related to refinable transition $T_4$ (see Figure 17).

TABLE III. $T_4$-HSP MAPPING

| Semantic Map | |
|---|---|
| ***T41*** | ***EnvInfoService*** |
| ***T42*** | ***PhyInfoService*** |
| ***T311*** | ***GeoInfoService*** |
| ***T432*** | ***TempInfoService*** |
| ***T441*** | ***SubjFeelAcqService*** |
| ***T442*** | ***EMRService*** |

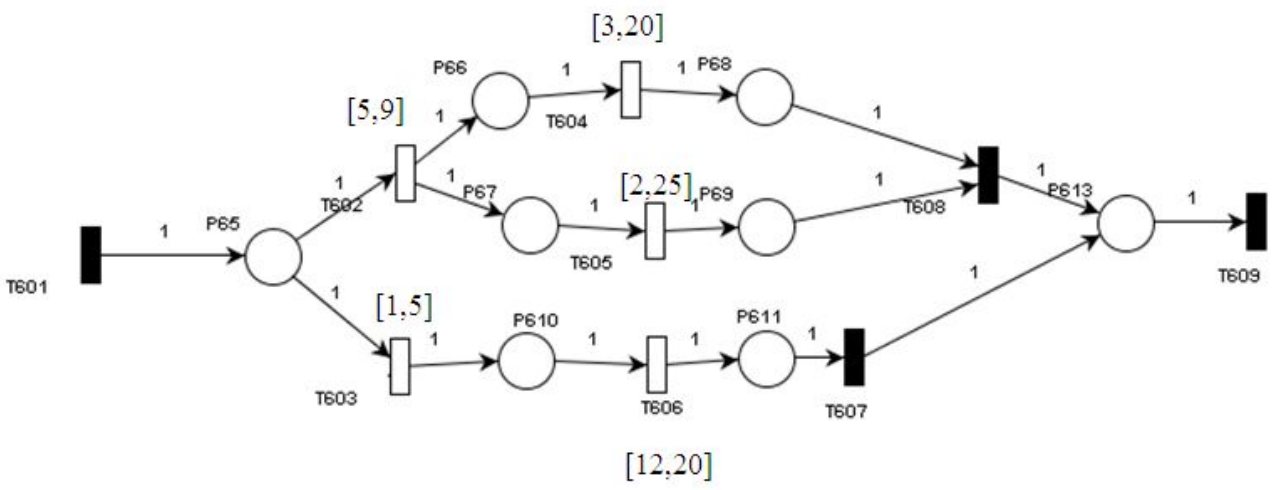

Figure 18. The model of refinable transition $T_6$

Table IV illustrates $T_6$–HSP mapping representing the model associated with refinable transition $T_6$ (see Figure 18).

TABLE IV. $T_6$-HSP MAPPING

| Semantic Map | |
|---|---|
| ***T602*** | ***HealthRiskAssService*** |
| ***T603*** | ***HealthGuideService*** |
| ***T604*** | ***RealTimeWarmService*** |
| ***T605*** | ***EmerAlarmService*** |
| ***T606*** | ***HealthGuideAssService*** |

The QoS calculation parameters are calculated and stored in order to be used by the optimal configuration selection algorithm. They also can be used to estimate or predict the QoS of composite service.

### B. *Experimentation*

To experiment the efficiency of our model, we developed a prototype of an Eclipse plug-in called *PetriNetWorkbench*. It aims to help the designer in building Petri-nets models for simulation purpose. Figure 13 shows an overview of four steps corresponding to phase of web services composition analysis implanted in this tool:

- *Domain analysis*: it concerns concepts identified in the textual description. These key concepts form a unified vocabulary of concepts that will be reusable for the description of user requirements.
- *Specification*: unified vocabulary that results from earlier step is used for specifying how both coherent and synthetic rules of user-requirements are.
- *Algorithms Generation*: automatic generation of detection algorithms from specification to check the validity of system model.
- *Detection*: this step gives the developed tool as input specifications and returns the conformity to user-requirement specification especially QoS attributes.

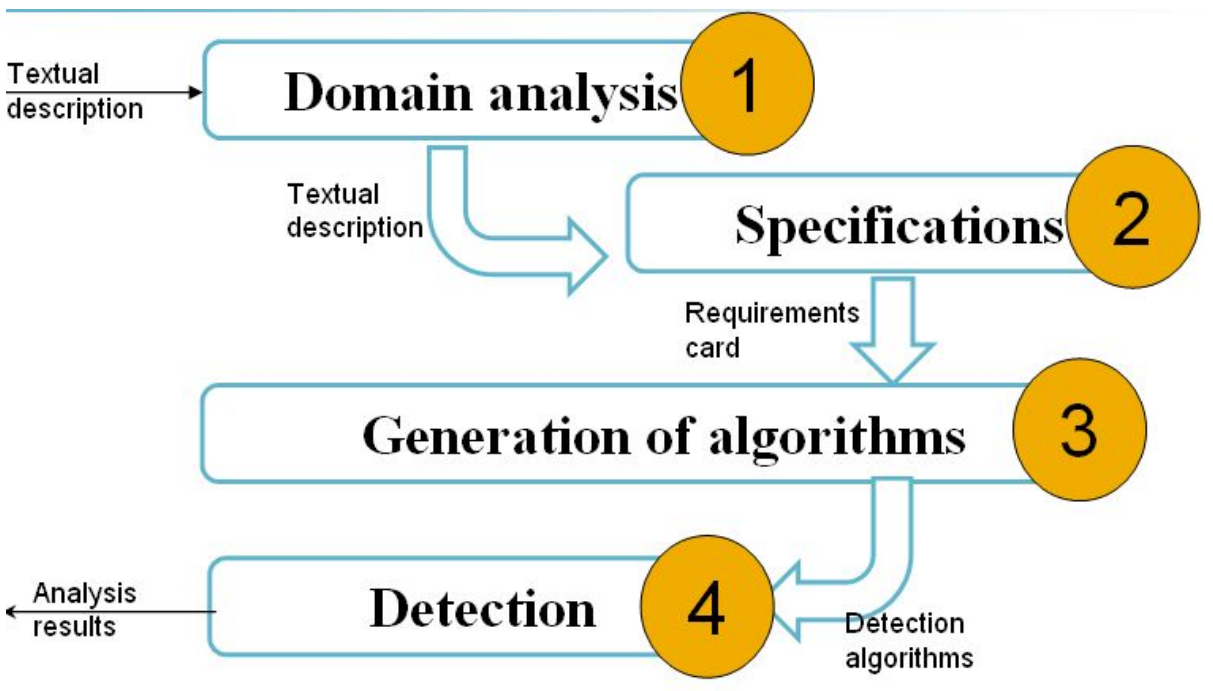

Figure 13. Analysis steps

Figure 20 depicts Petri-Net based model is defined by designer (only refinable transition $T_6$).

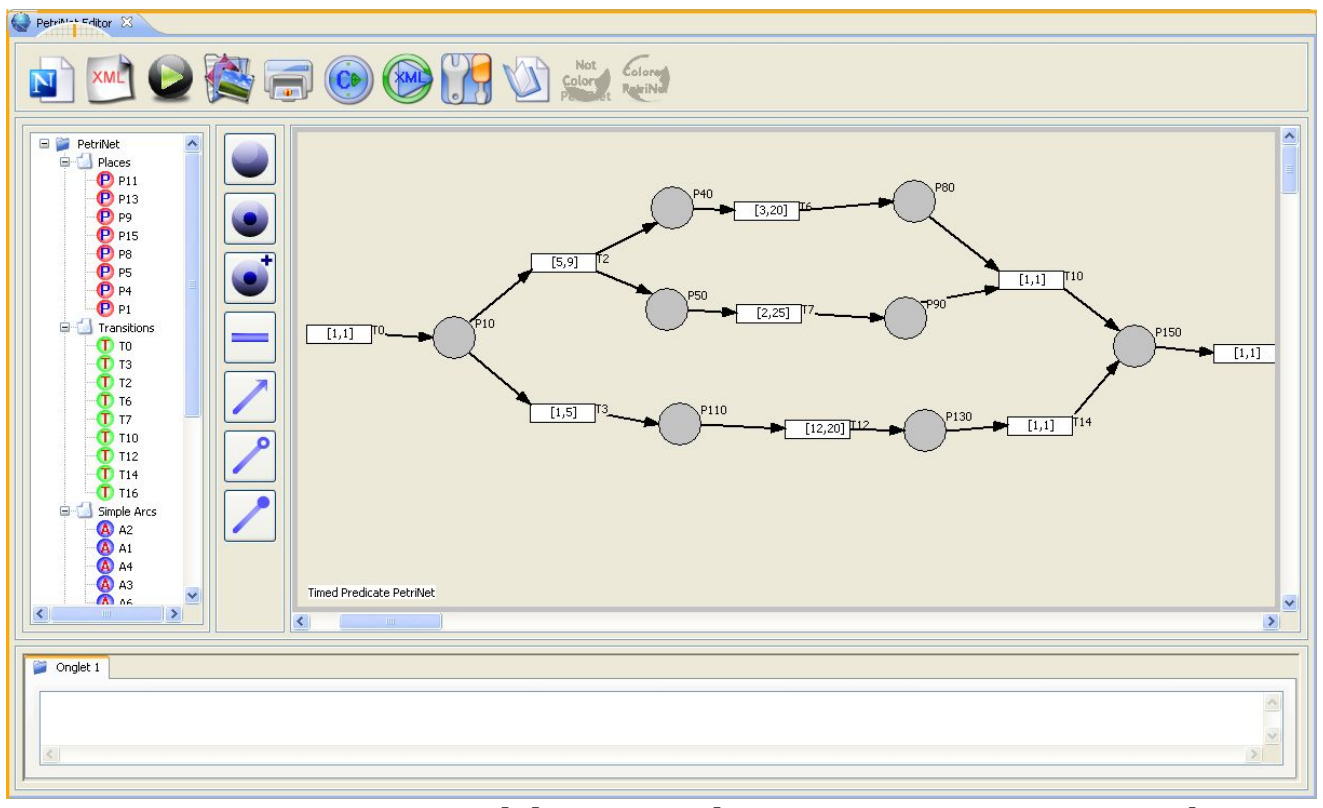

Figure 20. HTTPN model associated to transition $T_6$ creation with *PetriNetWorkbench*

The result of model analysis is returned after the simulation based on two steps: checking requirements card provided respecting a pre-defined grammar and then comparing the simulation result with the expected QoS requirements parameters defined by user.

## VII. CONCLUSION

In this paper, we presented a novel approach using high level Petri net with a rapid QoS calculation strategy. By Taking advantage of the global dynamic configuration, the composite service can adapt to the QoS dynamic change of components services and failure-prone runtime environment. The QoS requirements of users are satisfied to the greatest extent.

Further research is needed for sure. Firstly, the consideration of balanced configuration from different user QoS requirements is important in real deployment context of web services based applications. Secondly, we will extend the model to support dynamic QoS properties for future work. And finally, we will further the development of *PetriNetWorkbench* to support complex scenarios.

## AUTHORS PROFILE

**Sabri Mtibaa** is currently a Ph.D. student in the National School for Computer Sciences of Tunis, Tunisia (ENSI). He received the master degree from High School of Communication of Tunis, University of Carthage, Tunisia (Sup'Com) in 2008. His current research interest includes web service composition using Petri nets as well as system verification and QoS aware.

**Moncef Tagina** is a professor of Computer Science at the National School for Computer Sciences of Tunis, Tunisia (ENSI). He received the Ph.D. in Industrial Computer Science from Central School of Lille, France, in 1995. He heads research activities at LI3 Laboratory in Tunisia (Laboratoire d'Ingénierie Informatique Intelligente) on Metaheuristics, Diagnostic, Production, Scheduling and Robotics.